\documentclass{arxiv}

\usepackage{graphicx}
\usepackage{amssymb}
\usepackage{amsmath}
\usepackage{caption}
\usepackage{subfig}
\usepackage{cite}
\usepackage{multirow}
\usepackage{rotating}

\usepackage{epstopdf}

\begin{document}
\bibliographystyle{ieeetr}
\begin{frontmatter}

\title{Hardware Architecture for Large Parallel Array of Random Feature Extractors applied to Image Recognition}

\author[label1]{Aakash Patil\corauthref{cor1}},
\ead{aakashpatil@ntu.edu.sg}
\author{Shanlan Shen},
\author[label1]{Enyi Yao} and
\author[label1]{Arindam Basu\corauthref{cor1}}
\ead{arindam.basu@ntu.edu.sg}
\corauth[cor1]{Corresponding authors}
\address[label1]{School of Electrical and Electronic Engineering, Nanyang Technological University, Nanyang Avenue, Singapore 639798}

\begin{abstract}
We demonstrate a low-power and compact hardware implementation of Random Feature Extractor (RFE) core. With complex tasks like Image Recognition requiring a large set of features, we show how weight reuse technique can allow to virtually expand the random features available from RFE core. Further, we show how to avoid computation cost wasted for propagating ``incognizant" or redundant random features. For proof of concept, we validated our approach by using our RFE core as the first stage of Extreme Learning Machine (ELM)--a two layer neural network--and were able to achieve $>97\%$ accuracy on MNIST database of handwritten digits. ELM's first stage of RFE is done on an analog ASIC occupying $5$mm$\times5$mm area in $0.35\mu$m CMOS and consuming $5.95$ $\mu$J/classify while using $\approx 5000$ effective hidden neurons. The ELM second stage consisting of just adders can be implemented as digital circuit with estimated power consumption of $20.9$ nJ/classify. With a total energy consumption of only $5.97$ $\mu$J/classify, this low-power mixed signal ASIC can act as a co-processor in portable electronic gadgets with cameras.

\end{abstract}


\end{frontmatter}
\section{Introduction}
\label{Intro}

In recent years, image recognition capability has been improving with techniques like Convolutional Neural Network (CNN) and Deep Learning \cite{lee2014deeply, szegedy2014going,lin2014network}. These networks demand parallel computations of large set of feature extractors and are typically implemented on clusters of computers or graphics processing units (GPU). However, both these methods are area and power hungry--implementing image recognition on portable or wearable devices calls for customized low-power hardware implementations. One approach is to take inspiration from the low-power operation of the brain and design neural networks for pattern classification in a ``neuromorphic" way\cite{mitra_spikeclassify,true-north}. Proposed neuromorphic hardware solutions to date range from purely digital multi-core processor like SpiNNaker \cite{proceed2014Furber}, to active mode analog computing like HICANN \cite{schemmel2010wafer}, to sub-threshold analog approach like Neurogrid \cite{benjamin2014neurogrid}or floating gate technology like \cite{Lu_dml_isscc}. Similar to Neurogrid, we also use sub-threshold analog techniques for ultra low energy consumption; but with proper algorithm choice we are able to tolerate the process variation which are known to be major drawback for analog circuits\cite{kinget-mismatch}. Our hardware, on the other hand, exploits the process variations for realizing this randomness and can act as random feature extractor (RFE) core for certain layers of neural networks. As an example \cite{jarrett2009best, pinto2010evaluation, saxe2011random} observed that recognition accuracy with random convolutional filters is only slightly less than that of trained filters. Extreme Learning Machine (ELM) \cite{huang_elm_kernel} is another example that exploits randomness to realize its first stage without any training. As a proof of concept, in this paper we show how our hardware can be used as a RFE core to realize the first stage of ELM applied to image recognition.

There are several reported hardware architectures exploiting randomness in VLSI for ELM\cite{ijcnn-thakur,Patil_decode,chenyi_iscas2015}. Of these, \cite{ijcnn-thakur} shows the application of ELM to a single input single output regression problem. On the other hand, \cite{chenyi_iscas2015,Patil_decode} have already shown good accuracy at the system level for applications like intention decoding \cite{chenyi_iscas2015} and spike sorting \cite{Patil_decode} requiring multiple inputs and outputs--hence, we pursue this architecture further. The first novelty of this paper is in applying such a hardware to image based object recognition applications. With image pixels represented as digital values, we used digital input ELM (D-ELM) IC of \cite{Patil_decode} for our purpose. But, object recognition needs a  very large number of hidden neurons to achieve high accuracy--due to space constraints, this may be difficult to have in hardware. To circumvent this issue, we used a technique proposed in \cite{yao2015vlsi} for virtual expansion of the number of hidden neurons beyond the ones physically implemented on the chip. MATLAB simulations show that this method needs much fewer physical neurons to reach the accuracy levels obtained by using a large set of independent hidden neurons. However, due to uncontrolled randomness and systematic mismatch in hardware, not necessarily all neurons convey distinct information. To identify such redundant neurons, we propose a cognizance check as described in Section \ref{sec:cognizant}--this is the second novelty of this paper. The third novelty of this paper is the usage of simplified neuron models like tri-state ones which allow the usage of only adders in the second stage of ELM.

The organization of the paper is as follows: Section \ref{sec:algorithm} describes the network architecture of ELM and its training algorithms while  Section \ref{sec:hardware} shows the corresponding circuit blocks. Section \ref{sec:cognizant} describes the method for keeping cognizant neurons while removing redundant ones. We present results in Section \ref{sec:results} and conclude in the last section.

\begin{figure}
\centerline{
\includegraphics[width=0.8\textwidth]{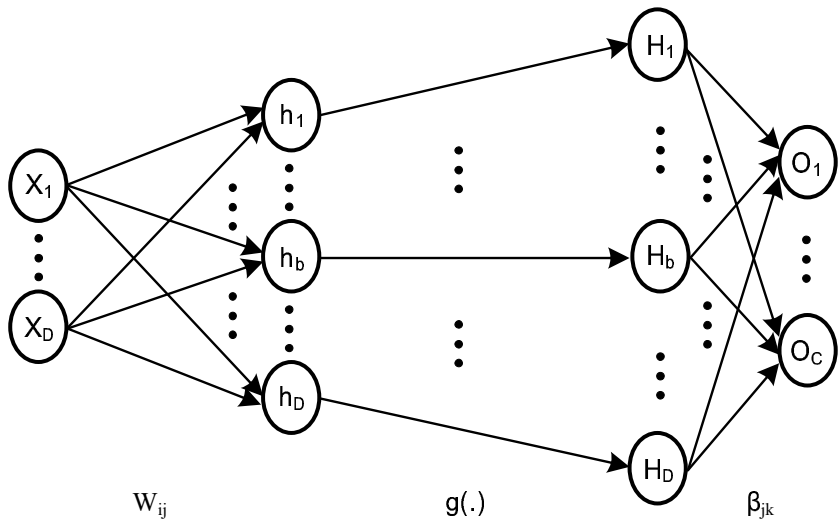}}
\caption{ELM network architecture where $x_j$ denotes j-th input dimension, $h_i$ is the input to the i-th hidden neuron obtained after random mixing of the inputs while $H_i$ denotes output of the i-th hidden neuron.} 
\label{fig:elm}
\end{figure}

\section{Machine Learning Algorithm}
\label{sec:algorithm}
\subsection{ELM Architecture}

For this work, we used the ELM algorithm with network architecture as shown in Fig. \ref{fig:elm}. It is a two stage neural network wherein the first stage maps the D-dimensional input vector $\textbf{X}$ nonlinearly to an L-dimensional hidden layer $\textbf{H}$ vector and the second stage linearly combines this hidden layer nodes to get $C$ output nodes. The output class predicted by ELM is the index of output node with highest value. The output value $O_k$ for $k^{th}$ node is given by:

\begin{equation}
O_k = \mathbf{H^T\beta_k} = \sum_{j=1}^{j=L}\beta_{kj}H_j
\end{equation}
where $H_j$ denotes the output of the j-th hidden neuron. This can be expressed as follows:
\begin{equation}
\begin{split}
H_j=g(h_j)=g(\mathbf{W_j^TX}+b_j)=g(\Sigma_{i=1}^{i=D}w_{ij}X_i+b_j)\\
\mathbf{W}_j,\mathbf{X}\in R^D; \mathbf{\beta_k,H} \in R^L; b_j,w_{ij}, \beta_{kj} \in R
\end{split}
\end{equation}
where $b_j$ is the bias for the $j^{th}$ hidden neuron, $w_{ij}$ are first stage weights, $\beta_{kj}$ are second stage weights and $g(.)$ is activation function of the hidden layer neuron.

The advantage of using ELM is that the $b_j$, $w_{ij}$ can be random numbers chosen from any continuous distribution while only $\beta_{kj}$ needs to be trained\cite{huang_elm}. Training $\beta_{kj}$ is also faster than iterative back-propagation methods with a closed form solution as given in \cite{huang_elm}: 
\begin{equation}
\label{eq:elm-train}
\beta=(\frac{I}{A}+\mathbf{ H^TH })^{-1}\mathbf{ H^TH }
\end{equation}
where $\textbf{H}$ is the hidden layer output and $\textbf{T}$ is the final expected output for the set of training samples. $A$ is the regularization factor to help generalization--it can be optimized by cross validation techniques\cite{huang_elm_kernel}. As per \cite{huang_elm}, many functions can be used for g(.). But for the ease of hardware implementation, we used two types of non-linearity: Rectified Linear Saturation Unit (RLSU) and tristate as shown below:
\begin{itemize}
\item Rectified Linear Saturated Unit (RLSU)
\begin{equation}
g(y)=
\begin{cases}
0 & \mbox{ for } y \leq 0 \\
y & \mbox{ for } 0 < y < th \\
th & \mbox{ for } y \geq th
\end{cases}
\end{equation}
\item Tristate
\begin{equation}
g(y)=
\begin{cases}
+1 & \mbox{ for } y \leq th \\
0 & \mbox{ for } -th < y < th \\
-1 & \mbox{ for } y \geq th
\end{cases}
\end{equation}
\end{itemize}
 With hidden layer outputs restricted to $+1$, $0$ and $-1$, resource costly digital multipliers of the second stage can be avoided in case of tristate non-linearity. The nonlinearity parameter th can be optimized by checking accuracy on validation samples.
 
\begin{figure}
\centerline{
\includegraphics[width=0.8\textwidth]{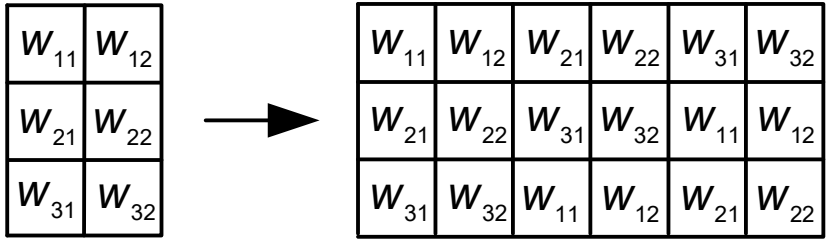}}
\caption{Diagram to explain the concept of hidden layer extension by weight rotation. Columns $3$ and $4$ are obtained by circularly shifting each row of the original weight matrix up by $1$ row.}
\label{fig:hidden_ext}
\end{figure}

\subsection{Virtual Expansion of Hidden Neurons}

\subsubsection{Technique of Virtual Expansion by Weight rotation}
An important parameter determining the classification capability of ELM is the number of hidden neurons $L$. In general, the accuracy improves with increasing number of hidden neurons. But implementing a large number of random weights and neurons will require larger chip area and power. So, we propose here a simple technique of reusing the available limited number of weight vectors and rotating them to get more ``virtual" weight vectors. For example, suppose the input-dimension for an application is $D$ and it requires $L$ hidden layer neurons. Conventionally, at least $D\times L$ random weights are needed for the random projection operation in the first layer of ELM to get the hidden layer matrix $\bf{H}$. However if the number of implemented hidden layer neurons is $N$ ($N < L$), the hardware can only provide a $D\times N$ random projection matrix $\bf{W}$ comprising weights $w_{ij}$(i = 1, 2, $\cdots$, D and j = 1, 2, $\cdots$, N). However, noting that we have a total of $D\times N$ random numbers on the chip, we can borrow concepts from combinatorics based learning\cite{shaista-neco,roy-lsmder,roy-tempo} to realize that the total number $N_w$ of $D$-dimensional weight vectors we can make is given by:
\begin{equation}
N_w = {D\times N \choose D}
\end{equation}
where the brackets denotes the operation of choosing $D$ unique items out of $D\times N$. For $N>>1$, $N_w$ grows almost as $D^N$. But the overhead of switches needed to allow all these combinations is prohibitively high. Instead, we choose a middle path and propose a method that creates $N_w'$ weight vectors ($N<N_w'<N_w$) but is easy to implement in hardware without needing additional switches.  A simple example of such an increased weight matrix is shown in Fig. \ref{fig:hidden_ext} for $D=3$ and $N=2$. This case shows the maximum increase possible to get a matrix of size $D\times (D\times N)$. Intuitively, each input dimension requires $L$ random numbers for the projection--it can be attained by reusing weights as long as $L<D\times N$. Next, we elaborate the method used to do this assuming $L<D\times N$.

To virtually expand the number of hidden layer neurons, we propose to do it in $\lceil{L/N}\rceil$ steps where the number of projections is increased $N$ in every step. For the second set of $N$ neurons, we need to shift the random matrix $\bf{W}$ comprising $w_{ij}$ (i = 1, 2,$\cdots$, D and j = 1, 2, $\cdots$, N) to $\bf{W}_{1,0}$ comprising $w_{ij}$ (i = 2, 3,$\cdots$, D, 1 and j = 1, 2, $\cdots$, N). Here, the subscript $(1,0)$ is used to denote a single circular rotation of the rows of the matrix $\bf{W}$. This notation implies $\bf{W}=\bf{W}_{0,0}=\bf{W}_{k,0}$. Using this notation, we can continue to get more random projections of the input (and thus expand the number of hidden neurons) by generating $\bf{W}_{1,0}$ to $\bf{W}_{\lceil{L/N}\rceil -1,0}$. To quantify this virtual expansion, we can define a virtual expansion factor $E$ as ratio of number of hidden neurons created ($L$) to number of independent random weight vectors available ($N$), i.e:
\begin{equation}
E=\frac{L}{N}
\end{equation}
This method is easily implemented in hardware since the input to the chip can be circularly rotated every time to effectively rotate the weights without adding any extra switches.
\begin{figure}
\centerline{
\includegraphics[width=0.8\textwidth]{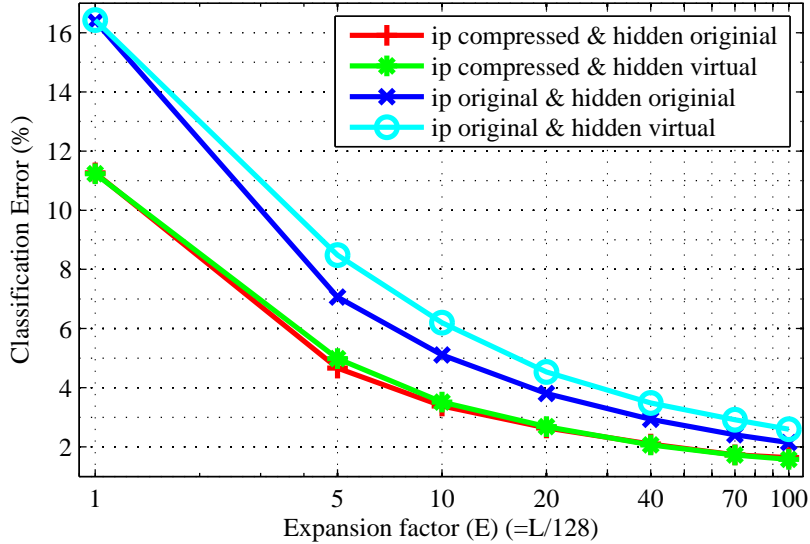}}
\caption{Classification error decreases with increasing L for all cases: `ip original' refers to original $784$ dimensional vector, `ip compressed' refers to $D=126$ dimensional vector obtained by $2\times3$ averaging, `hidden original' refers to independently generated $L=128E$ random weight vectors, `hidden virtual' refers to using only $128$ random weight vectors and rotating input to virtually expand by $E$ to create $L=128E$ hidden neurons.}
\label{fig:sw_rlsu}
\end{figure}

\subsubsection{Software modeling and validation of Weight Rotation}

We validated the technique of increasing the number of weight vectors by rotation using a software model in MATLAB. To model an independent set of log-normal weights due to mismatch of sub-threshold transistors\cite{chenyi_iscas2015,enyi-biocas}, we created a set of weights $w_{ij}=e^x$ where $x$ follows a gaussian distribution with $0$ mean and standard deviation of $0.6$. The reason for choosing this standard deviation is that the measured standard deviation of threshold voltage in this $0.35\mu$m CMOS process was $0.6U_T$ where $U_T$ denotes the thermal voltage $kT/q$. This model was simulated to get classification error for different value of $L=128*E$. For modeling our technique, we used just the first $128$ columns of this big matrix and rotated those vectors. Figure \ref{fig:sw_rlsu} shows that the classification error does decrease with increasing number of virtually created hidden neuron sets. This implies that those extra hidden neuron sets do provide extra information. In fact, the classification error obtained by using both ``independent neurons'' and ``virtual neurons'' have approximately the same classification ability. Also, to fit the dimension of $784$ for MNIST images within the $D=128$ input channels\cite{Patil_decode}, we applied a $2\times 3$ averaging of the image pixels to reduce the image dimension to $126$. From experimental results shown in Fig. \ref{fig:sw_rlsu}, we find that the compressed  ($2\times3$ averaging) MNIST image is easier to classify and consistently produces lower error than the original image. Hence, for all the experiments with hardware, we use $D=126$ pixel image.

\begin{figure}
\centerline{
\includegraphics[width=0.8\textwidth]{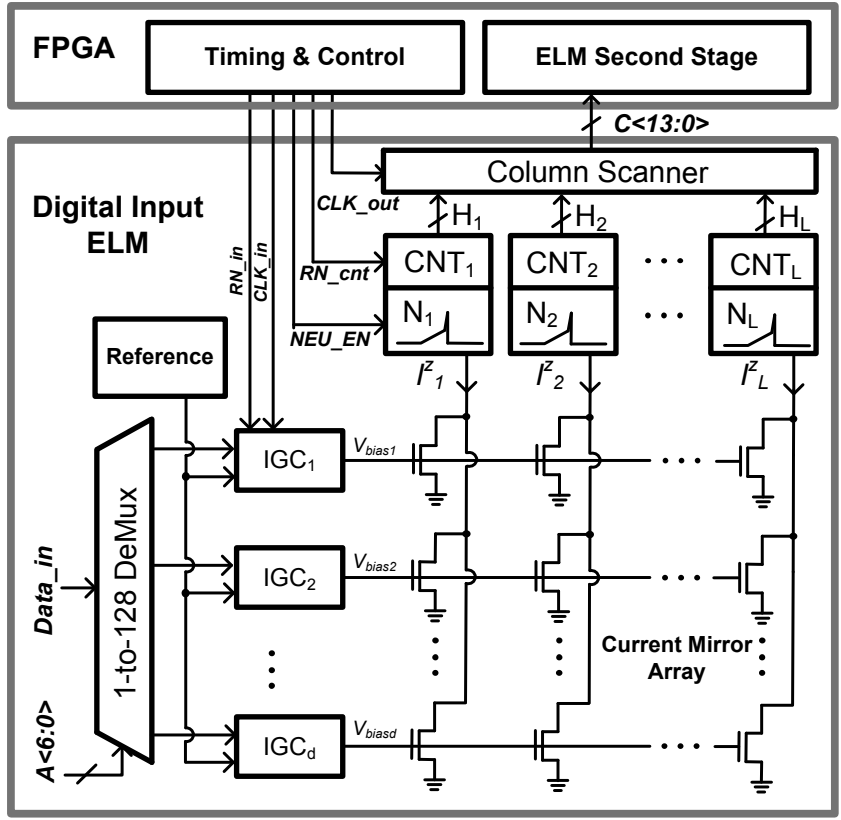}}
\caption{D-ELM IC architecture for implementing large parallel array of RFE and FPGA for performing virtual expansion and ELM second stage (input rotation is done in FPGA to virtually expand the number of hidden nodes).}
\label{fig:elm_archit}
\end{figure}

\begin{figure}[!t]
\centerline{
\subfloat[]{\includegraphics[width=0.3\textwidth]{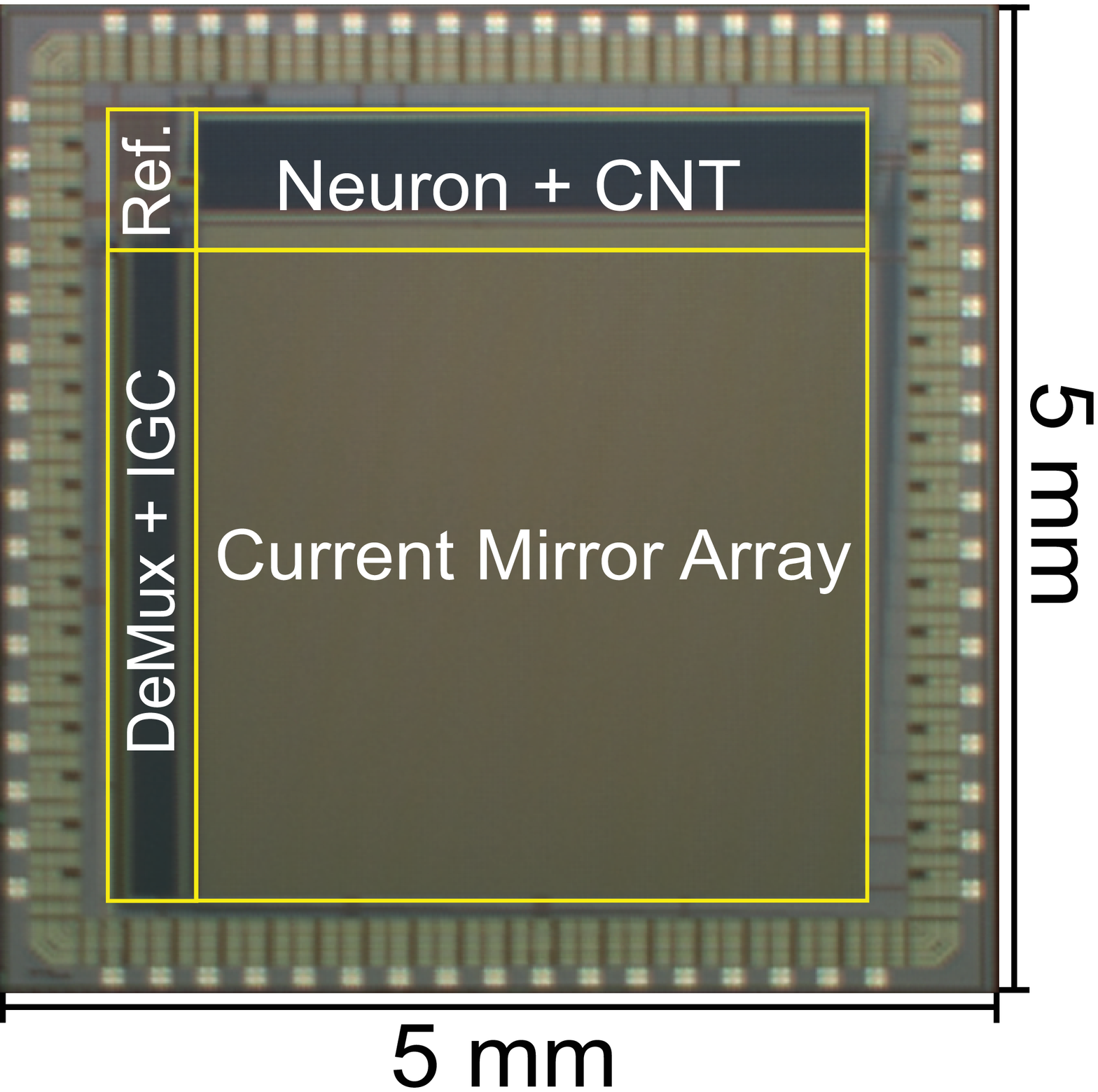}}
\subfloat[]{\includegraphics[width=0.4\textwidth]{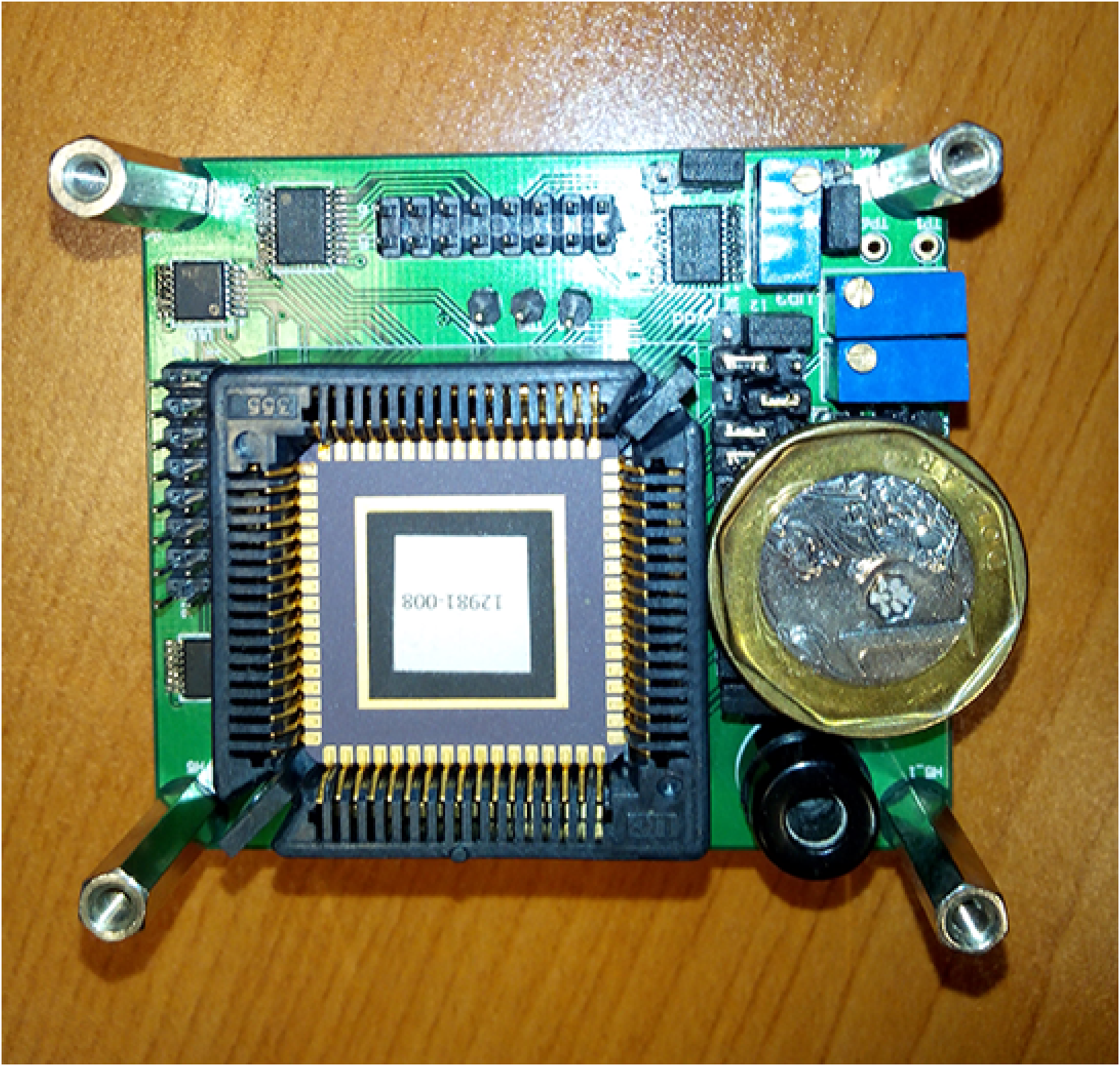}}}
\caption{(a) Die photo of the D-ELM IC fabricated in $0.35\mu$m CMOS measuring $5mm\times 5mm$ (b) Printed circuit board (PCB) designed for testing the IC.}
\label{fig:die_pcb}
\end{figure}

\section{Hardware Implementation}
\subsection{First stage: RFE core in analog ASIC}
\label{sec:hardware}
Figure \ref{fig:elm_archit} shows the architecture of the D-ELM consisting of three main parts: (1) Digital input $X_i$ is fed serially which is internally deserialized and fed to registers with specified addresses to create the input vector of dimension $D$ (2) input generation circuit (IGC) converts it into analog current which is scaled by random weights $w_{ij}$ of current mirror array (CMA) and added column-wise to perform the functionality of multiply and accumulate (MAC) (3) current controlled oscillator (CCO) converts the summed analog current into digital output $h_j$.

Random feature extraction functionality is performed by the CMA block consisting an array of $128\times 128$ current mirrors. Each mirror acts as a multiplier scaling the input current by a weight equal to ratio of driving strengths of input transistor and mirror transistor. The inherent mismatch in transistor driving strengths (owing primarily to threshold voltage mismatch in sub-threshold) enables us to create randomly distributed weights. With weights being defined by physical properties of transistors, it also acts as non-volatile storage of weights. Thus each single transistor of CMA acts as a multiplier cum weight storage element. Combining the output currents of mirrored transistors performs the adder functionality with no hardware cost. This compact realization of MAC and memory can enable to pack $128 \times 128$mirrors in a small area and enables up to $128\times 128$ MAC operations in parallel.

Fabricated in $0.35\mu m$ CMOS technology, the present D-ELM IC has a die area of $5mm\times 5mm$ (Fig. \ref{fig:die_pcb}(a)) with $128$ input registers and DAC occupying $0.7 mm \times 3.2 mm$ and the $128$ output CCO and counter occupying $3 mm \time 0.6 mm$. With minimum size transistors, the CMA array of $128 \times 128$ mirrors could have been realized in $0.1mm \times 0.12mm$ but presently occupies  $3.2 mm\times 3 mm$ to match the pitch of DAC and CCO along horizontal and vertical dimensions respectively. Figure \ref{fig:die_pcb}(b) shows the board designed to test the D-ELM chip.	 In the next version of this chip, we plan to utilize this unused area by including a bank of CMA. With functionality to connect one CCO to more than one mirror column, modified D-ELM IC can have much more hidden neurons in the same die area. Hence, in the modified D-ELM IC, we will rerun the IC using same input but can use the $\lambda^{th}$ CMA bank to get the $\lambda^{th}$ set of N hidden neurons. This CMA bank can also be used for increasing input dimension. For this we will have to rerun IC with $\lambda^{th}$ part of input and $\lambda^{th}$ CMA bank, but without resetting output counter so that outputs get accumulated as if a whole input vector is multiplied by weight vector. This can provide an  additional factor of hardware expansion over the rotation method used now. Even for applications requiring less number of input and hidden layer dimension, bank of CMA will help to select the most cognizant set of neurons described in the next section.

\subsection{Muting ``Incognizant" or Redundant Neurons}
\label{sec:cognizant}
Though process variations in IC technology enables to create a random weights with ease, but the uncontrollability of the same process variation in other parts of circuit can result in some neurons being biased towards always firing high (or low) independent of input. This can be caused by mismatch of the parameters of the neuron CCO as well as systematic variations. Hence, after the non-linearity, output of some neurons might saturate at small values of input i.e. they lack cognition to differentiate inputs over a wide range. Hence, they cannot propagate any information about input to later stages. Therefore, to save resources for later stages, it is better to ignore or power-down these redundant neurons. Reducing the $L$ hidden neurons to only $M$ ``cognitive" hidden neurons can help to reduce ELM’s training time by factor of $O$($M^2/L^2$) and reduce test resources by factor of $M/L$. For a given set of training samples, we count how many times a neuron output $H_j$ is saturating at positive/negative/zero threshold value. If this count is $>\theta\%$ ($\theta \approx 100$) of training samples, then we can assume that this neuron will give the same output for almost all of the test samples independent of its class. For software simulation with CMA modeled using lognormal random matrix in MATLAB, $M\backsim L$ is observed as seen in Table \ref{table:software}. We can see that this conclusion is valid independent of the type of neuronal nonlinearity. However, RFE done using D-ELM IC resulted in $M\backsim 0.7L$ for tristate and $M\backsim 0.8L$ for RLSU as shown later in Table \ref{table:hardware}. Lack of cognition in hardware might arise from systematic process variation or random process variations in CCO which scales the final output. If this scaling by CCO is very high (low), that neuron will always have high (low) value independent of scaling by the random weights of current mirrors. When the randomness of CCO is also modeled in MATLAB, we did observe $M<L$ confirming our suspicion.

\begin{table}[htbp]
\centering
\caption{\label{table:software}Simulated performance of the ELM for RLSU and Tristate neurons in software}
\begin{tabular}{|c|c|c|c|c|c|c|}
\hline
 & \multicolumn{3}{|c|}{RLSU} & \multicolumn{3}{|c|}{Tristate} \\
 \hline
 L & M & M/L & \%Error & M & M/L & \%Error \\
 \hline
 128 & 128 & 1.00 & 11.23 & 128 & 1.00 & 14.46 \\
  \hline
 640 & 637 & 1.00 & 4.98 & 637 & 1.00 & 6.34 \\
  \hline
 1280 & 1269 & 0.99 & 3.51 & 1274 & 1.00 & 4.46 \\
  \hline
 2560 & 2538 & 0.99 & 2.69 & 2547 & 0.99 & 3.41 \\
  \hline
 5120 & 5075 & 0.99 & 2.07 & 5084 & 0.99 & 2.68 \\
  \hline
 8960 & 8886 & 0.99 & 1.72 & 8896 & 0.99 & 2.34 \\
  \hline
 12800 & 12699 & 0.99 & 1.58 & 12730 & 0.99 & 2.10 \\
\hline
\end{tabular}
\end{table}

Our concept is inspired from a closely related concept of winner take all (WTA) and Principal Component Analysis (PCA), but still differs from them slightly. In WTA only the least active neurons (loser) are ignored; but there might be some greedy neurons which show high activity largely independent of the input. Our concept ignores not only the least active neurons (loser) but also the always highly active neurons (winner). One can term our concept as discriminator take all (DTA) where only neurons which discriminate inputs by showing different activity (winner for some and loser for some) for different inputs are propagated forward; somewhat like PCA which considers components with high variance. Another difference is that both PCA and WTA constrain the number of neurons to be selected based on ranking of their cognition capability, while our approach constrains cognition capability. In some cases WTA and PCA may result in selection of some neurons with low cognition capability; on other hand in some cases even neurons with good cognition capability are also ignored just because they do not rank in the group of highest cognition capability. Our approach does not rank neurons; rather, it just checks whether their cognition capability is good enough or not and mutes all those neurons with cognition capability below a certain limit, no matter how many of them are muted and how many selected.

One can easily point out more types of neurons which can be removed without loss of information. For example, neurons giving the same output value (even if not equal to threshold) are also incognizant. A neuron giving output correlated to some other neuron is also redundant. Calculating variance and co-variance of hidden layer output values for all training samples can give a good measure of cognition capability: more the variance more the cognition capability and less the co-variance lesser the redundancy. \cite{elm_pca} have reported use of PCA to reduce the number of hidden layer neurons of ELM. Their method may have given better selection of significant neurons, but would need additional transformation matrix to project original hidden layer neurons to additional layer of hidden layer neuron consisting only of principal components. Implementing this transformation matrix in hardware would require additional resources and hence we decided not to go with it.

\begin{figure}
\centerline{
\includegraphics[width=0.8\textwidth,trim={0 30 0 20},clip]{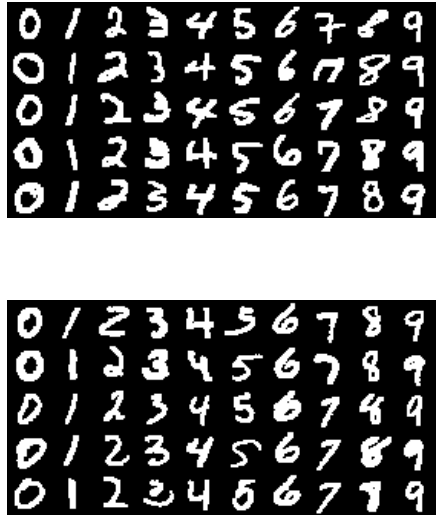}}
\caption{28 $\times$ 28 pixel images of MNIST samples: train samples (top) and test samples (bottom).}
\label{fig:mnist_sample}
\end{figure}

\section{Experiments and results}
\label{sec:results}
\subsection{Database}
For validating the performance of our hardware architecture, we used the MNIST \cite{mnist_cite} dataset of $70000$ images: training set of $60000$ images and test set of $10000$ images. Each sample data in MNIST is a 28 $\times$ 28 pixel grayscale image of handwritten digits. Figure \ref{fig:mnist_sample} shows sample images of handwritten digits from the MNIST database. Converted into a vector form, this would require an ELM with $D=784$. But as shown in Figure \ref{fig:sw_rlsu}, we found that compared to classification accuracy using original $784$ pixels, the accuracy is better by averaging neighboring pixel to convert it to 14 $\times$ 9 pixel image. This resulted in a $D=126$ dimensional input vector which was provided as input to D-ELM IC and its $N=128$ outputs were recorded. For increasing the hidden neurons, the IC was rerun $100$ times with the input vector circularly rotated. Thus for each of the image in MNIST, we created up to $12800$ random features.

\subsection{ELM training parameter optimization}
The random features for the training set were used to train the second stage weights according to equation (\ref{eq:elm-train}) wherein regularization factor `$A$' had to be optimized by cross-validation. The other parameter to be optimized was the threshold parameter `$th$'. This parameter optimization is expected to be data type dependent; hence, we can use a smaller training set to save the training time. We used $10000$ samples from training set to train the ELM with different `$A$' and `$th$' and then checked ELM accuracy for independent set of $20000$ validation samples from the training set. The parameter giving the best accuracy for this validation check was chosen for final training of ELM with full set of $60000$ training samples.

Training samples were also used to judge which $M$ of the $L$ random features are cognizant. With our dataset consisting of $10$ classes and almost equal number of samples for each class, chances are high that a given hidden neuron differentiates $9$ classes vs remaining class i.e. it has the same output value for $9$ classes ($\backsim 90\%$ samples). Hence, we need to have $\theta>90$. We used $\theta=99.5\%$ (corresponds to cognition only for $0.5\%$ of whole dataset or cognition for $\backsim 5\%$ of some class). As tabulated in Table \ref{table:hardware}, we observed $0.7X$ (for tristate) and $0.8X$ (for RLSU) reduction in the number of hidden neurons from the original set of hidden neurons. 

\begin{table}[htbp]
\centering
\caption{\label{table:hardware}Measured performance of the ELM for RLSU and Tristate neurons in hardware}
\begin{tabular}{|c|c|c|c|c|c|c|}
\hline
 & \multicolumn{3}{|c|}{RLSU} & \multicolumn{3}{|c|}{Tristate} \\
 \hline
 L & M & M/L & \%Error & M & M/L & \%Error \\
 \hline
 128 & 102 & 0.80 & 13.65 & 98 & 0.77 & 18.66 \\
  \hline
 640 & 511 & 0.80 & 7.85 & 474 & 0.74 & 8.32 \\
  \hline
 1280 & 1009 & 0.79 & 5.45 & 919 & 0.72 & 6.43 \\
  \hline
 2560 & 2068 & 0.81 & 4.01 & 1780 & 0.70 & 4.08 \\
  \hline
 3840 & 3180 & 0.81 & 2.96 & 2692 & 0.70 & 3.43 \\
  \hline
 5120 & 4196 & 0.81 & 3.01 & 3754 & 0.73 & 3.22 \\
  \hline
 6400 & 5235 & 0.82 & 2.75 & 4749 & 0.74 & 2.95 \\
  \hline
 7680 & 6340 & 0.83 & 2.62 & 5752 & 0.75 & 2.93 \\
  \hline
 8960 & 7378 & 0.82 & 2.45 & 6731 & 0.75 & 2.8 \\
  \hline
 10240 & 8502 & 0.83 & 2.52 & 7710 & 0.75 & 2.72 \\
  \hline
 11520 & 9498 & 0.82 & 2.27 & 8006 & 0.69 & 2.55 \\
  \hline
 12800 & 10314 & 0.81 & 2.14 & 9630 & 0.75 & 2.45 \\
\hline
\end{tabular}
\end{table}

Only the cognizant hidden neurons were used for training, reducing the training time for matrix inversion which now scales as $O(M^2) < O(L^2)$. An additional $L$-bit cognizance vector (with entries $1$'s and $0$'s for cognizant and incognizant hidden neuron respectively) will be needed to tell system which $M$ of $L$ neurons to be considered during testing phase reducing test resources which now scale as $O(M) < O(L)$. Using only $M < L$ hidden neurons, both first and second stage energy requirement reduces by factor of $O(M/L)$. With output weights quantized to $b_{\beta}$ bits, storing $L$ weights [$\beta_{k1}$ $\beta_{k2}$ $\beta_{k3}$ $\beta_{k4}$ .... $\beta_{kL}$] for each of $C$ output classes would have required $L$ $\times$ $C$ $\times$ $b_{\beta}$ bits of memory. But knowing only $M/L$ will be used we need only $L$ $\times$ $C$ $\times$ $b_{\beta}$ bits of memory for storing them (eg. say [$\beta_{k1}$ $\beta_{k4}$ .... $\beta_{kL}$]) reducing memory size by factor of $O(M/L)$. While performing second stage multiplications system can use cognizance vector (eg. [1 0 0 1 ... 1]), to determine whether to fetch or not fetch next $\beta_{kj}$ from memory reducing memory read operations by factor of $O(M/L)$.

\subsection{Second stage resource vs accuracy trade-off}
It can be seen from Table \ref{table:hardware} that the classification error keeps reducing with increasing number of virtual neurons reaching $2.14\%$ and $2.45\%$ for $L=12800$ RLSU and tristate neurons respectively. This classification errors were by using $6$bit weights $\beta$ in the output stage--this bit-width can be traded off as a resource for accuracy. As can be seen from Fig. \ref{fig:hw_rlsu_tristate}, the second stage has a resource vs accuracy trade-off that is outlined below:
\begin{figure}[!t]
\centerline{
\subfloat[]{\includegraphics[width=0.5\textwidth]{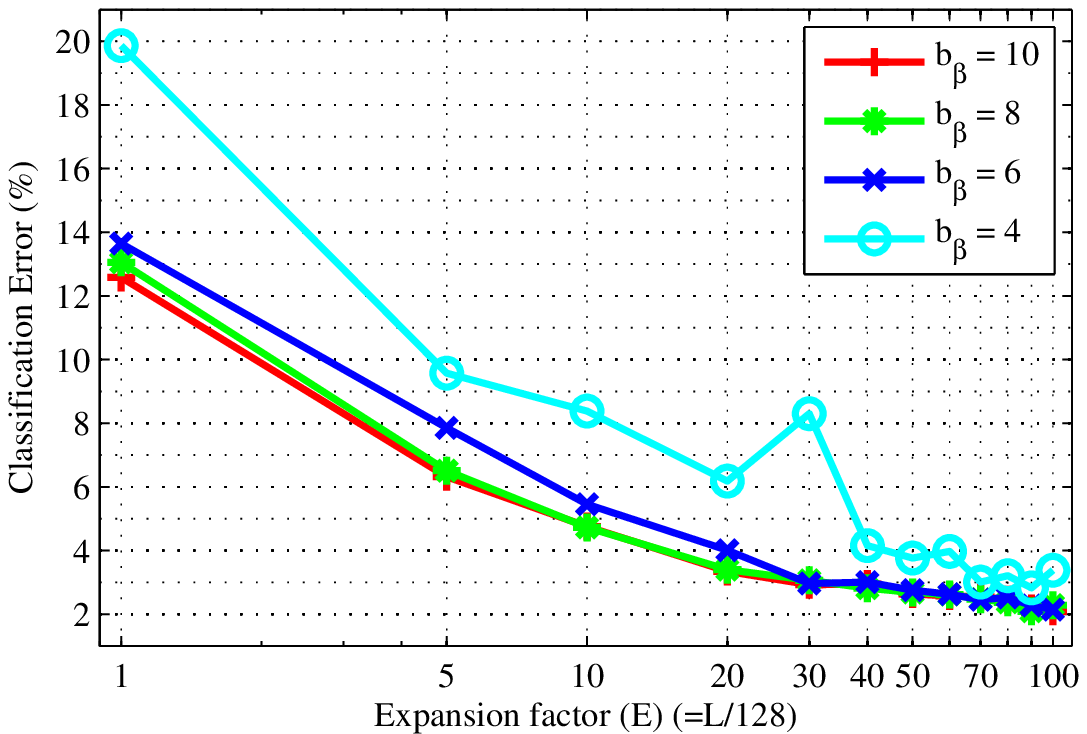}}
\subfloat[]{\includegraphics[width=0.5\textwidth]{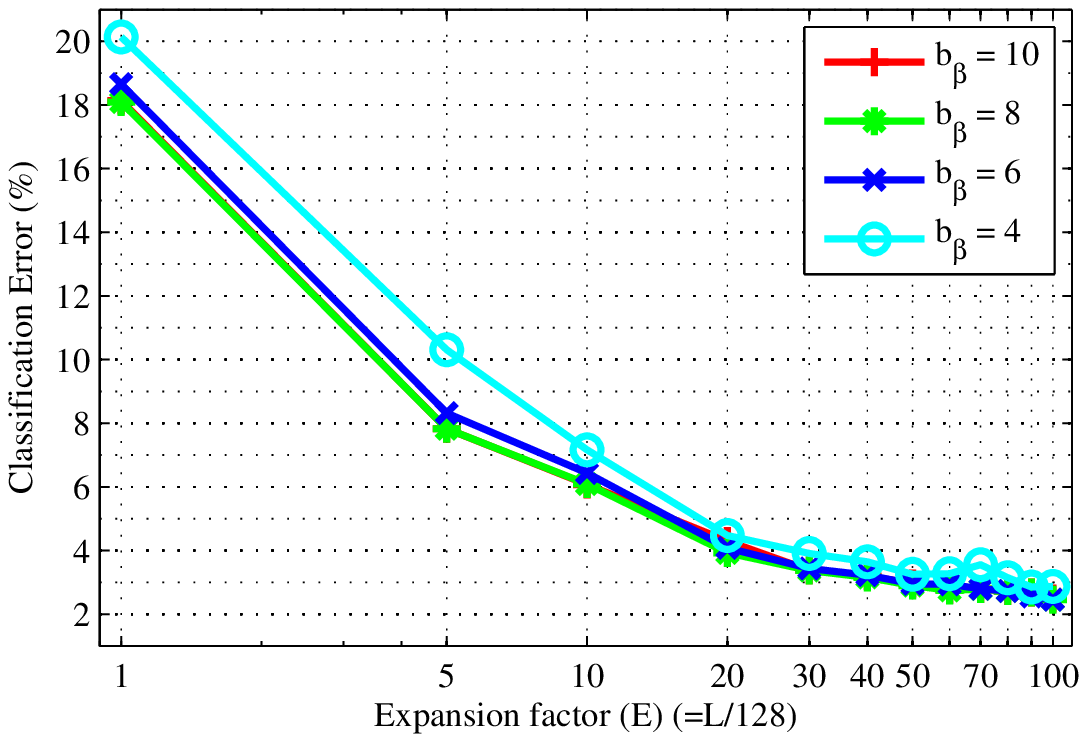}}}
\caption{Classification error decreases for increasing $L$ and increasing number of bits for output weights. (a) RLSU nonlinearity has slightly lower error than (b) tristate.}
\label{fig:hw_rlsu_tristate}
\end{figure}
\begin{itemize}
\item[(1)] Accuracy with RLSU nonlinearity is slightly better compared to that by tristate non-linearity; but use of tristate allows to eliminate multipliers. To get a better idea, we need to know the resource cost of MAC versus memory for the hardware platform. For a given accuracy, tristate nonlinearity may save on multipliers but may require higher $L$ and thereby higher memory for storing second stage weights. For RLSU nonlinearity, quantizing hidden neurons $H$ to lesser number of bits $b_H$ can help to reduce complexity of multiplier in MAC unit. We obtained D-ELM IC outputs $h$ to resolution of $b_h=12$, but due to saturation its output is limited to certain range and we quantized the output after RLSU to $b_H=8$ bits.
\item[(2)] Fast memory is very costly and quantizing output weights $\beta$ to lesser number of bits $b_{\beta}$ will help reduce the storage requirement. Lower $b_{\beta}$  also helps to reduce the complexity of multiplier in MAC unit. Decreasing $b_{\beta}$  did not affect accuracy much till a certain point. Quantizing second stage weights to $b_{\beta}=6$ (5 bits for magnitude and 1 bit for sign) was found to be good enough for both RLSU and tristate.
\item[(3)] Accuracy improves with increasing $L$ and $M$, but at the cost of more MAC (only accumulate in case of tristate nonlinearity) and more storage requirement for $\beta$. Accuracy improvement are diminishing especially for larger $L$. We could get accuracy of $97\%$ using $L=3840/6400$ for RLSU/tristate nonlinearity.
\end{itemize}
Even after increasing $L$ to very large value of $12800$, we found that classification accuracy cannot reach the accuracy levels of software implementation (accuracy $\backsim 98\%$ even for $L=5120$). One possible explanation is the extra noise in hardware (due to thermal/flicker noise of current mirrors and jitter in CCO) which is currently not modeled in MATLAB. Another reason might be the systematic randomness in hardware which results in less number of cognizant neurons.

\subsection{Energy Efficiency}
From characterization results (refer \cite{yao2015vlsi} for details), the RFE-core consumes $188.8$ $\mu$W from $1$ V power supply while operating at a speed of $31.6$K random projections/second/CCO with $D=128$, $L=100$ . From this value, we can estimate the energy needed to perform classifications in our case. For $D=126$, $N=128$ and $2000$ random projections/second/CCO, we can estimate the total energy consumption of $\approx 119$ nJ/conversion where a conversion refers to $128$ random projections happening in parallel. For an example case of $L=6400$, we need to rerun each image $50$ times that results in $5.95$ $\mu$J/classify at a speed of $40$images/sec. From training, we know only $M=4749$ neurons are cognizant and hence while testing, we do not waste resource for further operations on them. To create zero-mean random vectors, we need to perform $4749$ pairwise subtractions in a method similar to \cite{Patil_decode}. Since we plan to use tristate non-linearity, ELM 2nd stage is just 4749$\times$10 addition/subtraction operation of $10$ sets of output weights. Energy efficient architectures \cite{reynders2011} can enable implementation of $32$ bit accumulate at $0.4$ pJ/addition resulting in second stage energy consumption of $20.9$ nJ/image. So if a dedicated digital circuit implementing the second stage is integrated with D-ELM, the net energy consumption will be $5.97$ $\mu$J/classify. We can also report a more conventional metric of energy/MAC--for our system, this is dominated by the CCO and is equal to $\approx 7$ pJ/MAC.

Table \ref{table:comparison} shows the energy/classify for other hardware implementations. Our hardware implementation has lesser error rate than \cite{minitaur,spinnaker_mnist} but still consumes orders of magnitude less energy than them. Using back-propogation to train the network TrueNorth\cite{truenorth_mnist} is able to achieve much lesser error rate of 0.58\% but will require high energy consumption of $108$ $\mu$J/classify. For lower energy configuration their error rate increases and with comparable error rate of 5\% our hardware will have lesser energy consumption than them.

\begin{table}[htbp]
\centering
\caption{\label{table:comparison}Comparison of works performing MNIST classification on hardware}
\begin{tabular}{|c|c|c|c|}
\hline
 hardware & design approach & \%error & energy/classify\\
 \hline
 Minitaur\cite{minitaur} & FPGA & 8\% & $200$ mJ/classify \\
 \hline
 SpiNNaker\cite{spinnaker_mnist} & multi ARM core & 5\% & $6$ mJ/classify \\
 \hline
 \multirow{3}{*}{TrueNorth\cite{truenorth_mnist}} & \multirow{3}{*}{custom digital} & 0.58\% & $108$ $\mu$J/classify \\
 & & 5\% & $4$ $\mu$J/classify \\
 & & 7.3\% & $268$ nJ/classify \\
 \hline
 \multirow{3}{*}{D-ELM [this]} & \multirow{3}{*}{custom mixed-signal} & 2.95\% & $6$ $\mu$J/classify \\
 & & 4.08\% & $2.4$ $\mu$J/classify \\
 & & 3.43\% & $3.6$ $\mu$J/classify \\
\hline
\end{tabular}
\end{table}

\section{Conclusion and Future Work}
\label{sec:conclusion}
We present a proof of concept of how analog sub-threshold techniques can be used for realization of large parallel array of Random Feature Extractors for image recognition hardware with ultra-low energy consumption suitable for portable electronic gadgets. We propose a weight reuse technique for virtually increasing the random features available from hardware with limited output features. Further we also show how cognition check can enable to mute the incognizant feature extractors in hardware. In future, we plan to have a much bigger array of synapse weights which will provide more options to selectively choose most cognizant neurons.

Implementing fully connected two stage ELM, our hardware approach for was able to achieve $> 97\%$ accuracy on MNIST. In future we wish to validate our approach for complex dataset like NORB and CIFAR. Another direction we are exploring is using our IC as random convolutional filter as shown in \cite{jarrett2009best, pinto2010evaluation, saxe2011random}; however, we will need to rerun our IC many times for convolution around each pixel. An alternative can be using our IC for receptive field based ELM (RF-ELM) \cite{mcdonnell2015plosone}, but will need to think of easy ways to turn on/off patches of current mirrors in our CMA block without much area and power overhead. 

\section{Acknowledgements}
Financial support from MOE through grants RG 21/10 and ARC 8/13 and from SMART Innovation grant are acknowledged.
\bibliography{elm_hardware_aakash-basu}

\end{document}